\documentclass[aps, prc, floatfix, twocolumn, showpacs, preprintnumbers, superscriptaddress, nofootinbib]{revtex4-1}

\usepackage[10pt]{type1ec}  
\usepackage[T1]{fontenc}
\usepackage{CJKutf8}
\usepackage[overlap, CJK]{ruby}
\usepackage{CJKulem}
\newenvironment{SChinese}{%
  \CJKfamily{gbsn}%
  \CJKtilde
  \CJKnospace}{}

\usepackage{graphicx}
\usepackage{amssymb,amsmath}
\usepackage{color}
\usepackage{isotope}
\usepackage{xspace}
\usepackage{isotope}
\usepackage[colorlinks=true,citecolor=blue,urlcolor=blue]{hyperref}

\pdfoutput=1
\graphicspath{{figs/}}

\newcommand{\csq}{c^2_{s}}
\newcommand{\csqmax}{c^2_{s, {\rm max}}}

\newcommand{\mincsq}{\textbf{min}\{c^2_{s, {\rm max}}\}}
\newcommand{\csqmat}{c^2_{s, {\rm match}}}
\newcommand{\nb}{n_{\rm B}}
\newcommand{\nsat}{n_{\rm sat}}
\newcommand{\Hz}{\, \text{Hz}}
\newcommand{\km}{\, \text{km}}

\newcommand{\fmiq}{\, \text{fm}^{-3}}
\newcommand{\MeV}{\, \text{MeV}}

\newcommand{\La}{\Lambda}

\newcommand{\De}{\Delta}
\newcommand{\ep}{\varepsilon}

\newcommand{\emat}{\varepsilon_{\rm m}}
\newcommand{\Mmax}{M_{\rm max}}
\newcommand{\Msun}{{\rm M}_{\odot}}
\newcommand{\Mchirp}{{\mathcal M}}

\newcommand{\Rtwo}{R_{2.0}}

\newcommand{\NNNLO}{\ensuremath{{\rm N}{}^3{\rm LO}}\xspace}
\newcommand{\NNLO}{\ensuremath{{\rm N}{}^2{\rm LO}}\xspace}
\newcommand{\chiEFT}{$\chi$EFT\xspace}

\newcommand{\article}{article\xspace}
\newcommand{\secref}[1]{Sec.~\ref{#1}\xspace}

\newcommand{\eg}{\textit{e.g.}\xspace}
\newcommand{\ie}{\textit{i.e.}\xspace}
\newcommand{\etal}{\textit{et~al.}\xspace}

\definecolor{sr}{rgb}{0.51, 0.67, 0.7}
\definecolor{cd}{rgb}{0., 0.5, 0.9}

\allowdisplaybreaks

\begin{document}
\preprint{INT-PUB-21-018, N3AS-21-011}
\title{Large and massive neutron stars: Implications for the sound speed in dense QCD}

\author{Christian~Drischler}
\email{drischler@frib.msu.edu}
\affiliation{Facility for Rare Isotope Beams, Michigan State University, East Lansing, MI~48824, USA}

\author{Sophia~Han 
(\begin{CJK}{UTF8}{}\begin{SChinese}韩 君\end{SChinese}\end{CJK})
}
\email{sjhan@berkeley.edu}
\affiliation{Institute for Nuclear Theory, University of Washington, Seattle, WA~98195, USA}
\affiliation{Department of Physics, University of California, Berkeley, CA~94720, USA}

\author{Sanjay~Reddy}
\email{sareddy@uw.edu}
\affiliation{Institute for Nuclear Theory, University of Washington, Seattle, WA~98195, USA}

\date{October 28, 2021}

\begin{abstract}

The NASA telescope {\it NICER} has recently measured x-ray emissions from the heaviest of the precisely known two-solar mass neutron stars, PSR J0740+6620. Analysis of the data  [Miller~\textit{et~al.},  Astrophys. J. Lett. 918, \textbf{L28} (2021); Riley~\textit{et~al.}, Astrophys. J. Lett. 918, \textbf{L27} (2021)] suggests that PSR J0740+6620 has a radius in the range of $R_{2.0} \approx (11.4-16.1) \, \text{km}$ at the $68\%$ credibility level. In this article, we study the implications of this analysis for the sound speed in the high-density inner cores by using recent chiral effective field theory~($\chi$EFT) calculations of the equation of state at next-to-next-to-next-to-leading order to describe outer regions of the star at modest density. 
We find that the lower bound on the maximum speed of sound in the inner core, $\mincsq$, increases rapidly with the radius of massive neutron stars. If $\chi$EFT remains an efficient expansion for nuclear interactions up to about twice the nuclear saturation density, $R_{2.0}\geqslant 13$ km requires $\mincsq \geqslant 0.562$ and $0.442$ at the 68\% and 95\% credibility level, respectively.  

\end{abstract}

\maketitle

\section{Introduction}

The dedicated NASA x-ray telescope {\it {Neutron Star Interior Composition ExploreR}} ({\it NICER}) has recently measured soft x-ray emissions from hotspots on the surface of the millisecond pulsar PSR J0740+6620---the heaviest of the precisely known two-solar mass neutron stars, $M=2.08\pm 0.07\,\Msun$~\cite{Cromartie:2019kug,Fonseca:2021wxt}. 
Independent analyses of the {\it NICER} and {\it XMM-Newton} data combined suggest that PSR J0740+6620 also has a large radius in the range of $\approx (11.4-16.1) \km$ at the $68\%$ credibility level. 
Specifically, the Maryland-Illinois group inferred $(12.2-16.1)\km$~\cite{Miller:2021qha}, while the x-ray Pulse Simulation and Inference (XPSI) group obtained with $(11.4-13.7)\km$~\cite{Riley:2021pdl} statistically consistent but somewhat smaller radii.\footnote{
The XPSI group~\cite{Riley:2021pdl} used a larger calibration uncertainty compared to the Maryland-Illinois group~\cite{Miller:2021qha} which permits lower inferred radii, and also a hard upper limit (inferred from theoretical nuclear models) such that the prior support is zero for radii greater than $16\km$.} 
Radii $>13 \km$ for $\approx 2\,\Msun$ stars would have profound implications for the properties of the dense matter equation of state (EOS) in neutron star cores as well as the composition and phases of dense quantum chromodynamics (QCD) at low temperatures. 
In this \article, we explore some of these implications for the sound speed of large and massive neutron stars.

Astronomical observations can provide important constraints on the EOS since neutron star properties such as masses, radii, and tidal deformabilities are sensitive to the EOS in the (baryon) density regime $\nb \approx (2-4)\,\nsat$. Here, $\nsat = 0.16\fmiq$ denotes the canonical value for the empirical nuclear saturation density, a typical density in heavy atomic nuclei. In particular, radio observations have provided precise mass measurements of three neutron stars with masses $\approx 2\,\Msun$~\cite{Demorest:2010bx,Antoniadis:2013pzd,Cromartie:2019kug,Fonseca:2021wxt}. Such high masses require high matter pressures in the neutron star core, disfavoring strong first-order phase transitions in this density range. Further, the first direct gravitational wave (GW) detection from the binary neutron star merger GW170817 provided stringent constraints on the tidal deformability of neutron stars with canonical masses $\approx 1.4\,\Msun$~\cite{Abbott:2018exr, De:2018uhw,Capano:2019eae}. The firm upper bound on the tidal deformability inferred from GW170817 has indicated a relatively small radius $\lesssim 13.4\km$~\cite{Abbott:2018exr, Landry:2020vaw,Al-Mamun:2020vzu} for $\approx 1.4\,\Msun$ stars, suggesting that the pressure of matter at the densities encountered in the 
outer core is low. 
Analysis of x-ray observations of surface thermal emissions from quiescent neutron stars in low-mass x-ray binaries (LMXBs) also favor small radii in the range $(10-12) \km$~\cite{Ozel:2016oaf,Bogdanov:2016nle}, although the systematic uncertainties are still large~\cite{Steiner:2015aea,Steiner:2017vmg}. 
When combined, radio and GW observations favor a rapid transition from low to high pressures toward the inner core. 
In natural units this corresponds to sound speeds $c_s\geqslant \sqrt{1/3}$ because $\csq =\partial P(\ep)/\partial \ep$ is the derivative of the pressure $P(\ep)$ with respect to the energy density $\ep$ including rest mass contributions~\cite{Bedaque:2014sqa,Tews:2018kmu}.

The purpose of this \article is to determine the minimum sound speed in the neutron star core required to support radii in the range of $\Rtwo \approx (11.4-16.1)\km$, where $\Rtwo$ is the radius of a neutron star with mass $M=2.0\,\Msun$. To this end, we use recent microscopic EOS constraints~\cite{Drischler:2017wtt,Leonhardt:2019fua,Drischler:2020hwi,Drischler:2020yad} derived from chiral Effective Field Theory (\chiEFT) up to next-to-next-to-next-to-leading order (\NNNLO) 
with correlated \chiEFT truncation errors~\cite{Melendez:2019izc} quantified. These errors, which arise due to truncating the \chiEFT expansion at a finite order in practice, are important to quantify as they can be significant for $\nb \gtrsim \nsat$, 
even at \NNNLO in the \chiEFT expansion (see, \eg, Ref.~\cite{Drischler:2021kxf} for a recent review article).

This remaining \article is organized as follows. 
We discuss in Sec.~\ref{sec:outer_core} the importance of theoretical calculations and experimental constraints on the nuclear matter EOS in the density region of $(1-2)\,\nsat$. Section~\ref{sec:inner_core} assesses how the current uncertainties in microscopic EOS calculations up to $\sim2\,\nsat$ impact the interpretation of neutron star observations about the sound speeds of high-density matter. 
In \secref{sec:cons}, we conclude by discussing the implications of high sound speeds and large neutron star radii for dense matter physics and multi-messenger astronomy. 
We use natural units in which $\hbar = c = 1$.

\section{Equation of state of the outer core} 
\label{sec:outer_core}

In a recent article~\cite{Drischler:2020fvz}, we showed how upper and lower bounds on the neutron star radius of any mass can be derived from microscopic EOS calculations up to some density $n_c$; typically, $n_c\lesssim 2.0\,\nsat$. 
Matching these microscopic calculations at $\nb=n_c$ to a maximally stiff EOS characterized by a constant sound speed $\csqmat\leqslant 1$ for $\nb \gtrsim n_c$ allows one to obtain robust upper and lower bounds on the neutron star radius. 
If the pressure and energy density are matched continuously at $\nb=n_c$, the maximum radius  associated with that EOS can be inferred; whereas matching with a maximal finite discontinuity in the energy density specified by an assumed lower bound on $\Mmax$ determines the minimum radius. 
Here, we extend this work to provide a lower bound on the maximum speed of sound, $\mincsq$, reached in PSR J0740+6620's core for a given radius $\Rtwo$ in the range of the recent {\it NICER} measurement.

\begin{figure}[tb]
\begin{centering}
\includegraphics[scale=1]{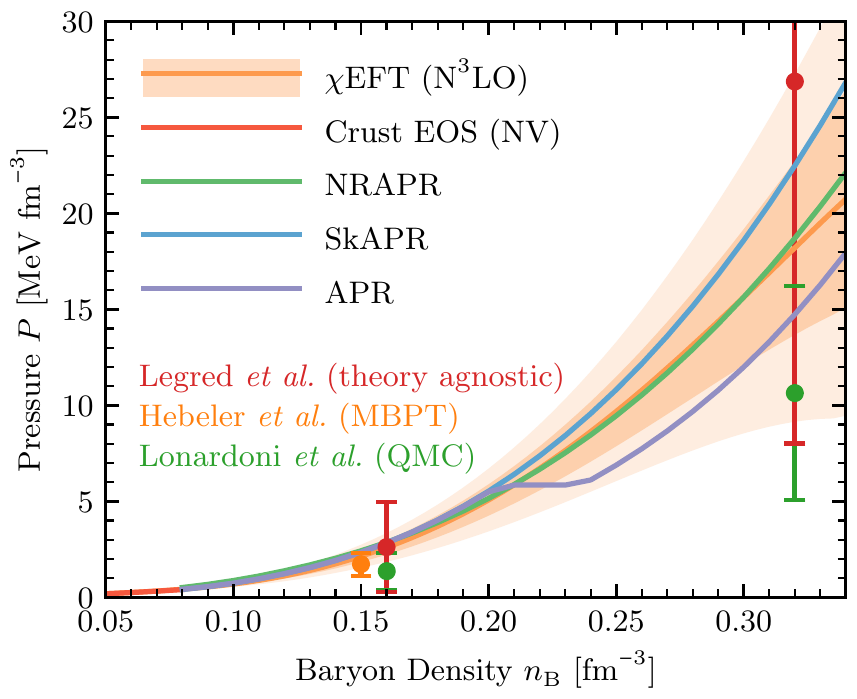}
\end{centering}
\caption{
The pressure $P$ 
as a function of the baryon density $\nb$ in neutron star matter. The uncertainty bands 
(dark: $1\sigma$, light: $2\sigma$) were derived in Ref.~\cite{Drischler:2020fvz} using the \NNNLO results obtained in Refs.~\cite{Drischler:2020hwi, Drischler:2020yad} for the interaction with the momentum cutoff $500\MeV$ [``GP--B~$(500\MeV)$''].
The error bars 
show the microscopic constraints from Hebeler~\etal~\cite{Hebeler:2013nza} (based on lower-order MBPT calculations) and Lonardoni~\etal~\cite{Lonardoni:2019ypg} (based on QMC calculations) as well as the theory-agnostic constraints from neutron star observation by Legred~\etal~\cite{Legred:2021hdx}. The green, blue, and purple lines depict the phenomenological EOSs NRAPR, SkAPR, and APR in beta equilibrium~\cite{Constantinou:2014hha,Schneider:2019vdm}, respectively. Note that the APR EOS involves a first-order transition into a pion condensate around $1.3\,\nsat$. 
At $\nb \leqslant 0.08 \fmiq$, we also show the crust EOS by Negele \& Vautherin~\cite{Negele:1971vb} (NV, red line). }
\label{fig:pressure_nsm}
\end{figure}

Following the strategy discussed in Ref.~\cite{Drischler:2020fvz}, we construct the neutron star EOS from low to high densities by matching EOSs defined in three different density regions. At low densities, $\nb \leqslant 0.5\,\nsat$, we use the standard crust EOS derived by Baym, Pethick, and Sutherland~\cite{Baym:1971pw} and Negele \& Vautherin~\cite{Negele:1971vb}. At intermediate densities, $0.5\,\nsat < \nb \leqslant n_c$, we interpolate microscopic calculations of the EOS in pure neutron matter (PNM) and symmetric nuclear matter (SNM) to beta-equilibrated matter, \ie, neutron star matter (NSM), using the standard quadratic expansion of the EOS's isospin dependence. 
Explicit calculations of isospin asymmetric matter based on \chiEFT nucleon-nucleon (NN) and three-nucleon (3N) interactions have shown that the standard quadratic expansion is a reasonable approximation~\cite{Drischler:2013iza, Drischler:2015eba, Kaiser:2015qia, Wellenhofer:2016lnl, Somasundaram:2020chb, Wen:2020nqs}. We refer to this region as the outer core.

Specifically, in the outer core, we consider the microscopic constraints on the zero-temperature EOS in PNM and SNM obtained in Refs.~\cite{Drischler:2017wtt, Leonhardt:2019fua, Drischler:2020hwi} by high-order many-body perturbation theory (MBPT) calculations. 
These many-body calculations with nonlocal NN and 3N interactions at the same order in the \chiEFT expansion have improved previous state-of-the-art MBPT results~\cite{Tews:2012fj, Kruger:2013kua, Drischler:2016djf}, allow for estimating EFT truncation errors up to \NNNLO, and were recently used in the first statistical analysis of correlated EFT truncation errors in nuclear matter~\cite{Drischler:2020hwi,Drischler:2020yad}. Further, the underlying \NNLO and \NNNLO interactions exhibit reasonable nuclear saturation properties in SNM [as the leading short-range 3N forces were adjusted to the empirical saturation point] and predict the nuclear symmetry energy evaluated at $\nsat$ and its slope parameter in excellent agreement with experimental constraints (see Figure~2 in Ref.~\cite{Drischler:2020hwi}).
Similar to other \chiEFT calculations, the slope parameter is predicted significantly lower than the mean value of the recent PREX--II-informed constraint obtained from covariant energy density functionals~\cite{Reed:2021nqk}, $L=(106 \pm 37)\MeV$. However, the microscopic EOS used here is consistent at the 68\% level with the PREX--II-informed constraint due to the large experimental uncertainties. It should also be noted that the PREX--II measurement of the \isotope[208]{Pb} neutron skin is in tension with constraints from the \isotope[208]{Pb} dipole polarizability, which has not yet been reconciled~\cite{Reinhard:2021utv,Piekarewicz:2021jte}.

\begin{figure}[tb]
\begin{centering}
\includegraphics[scale=1]{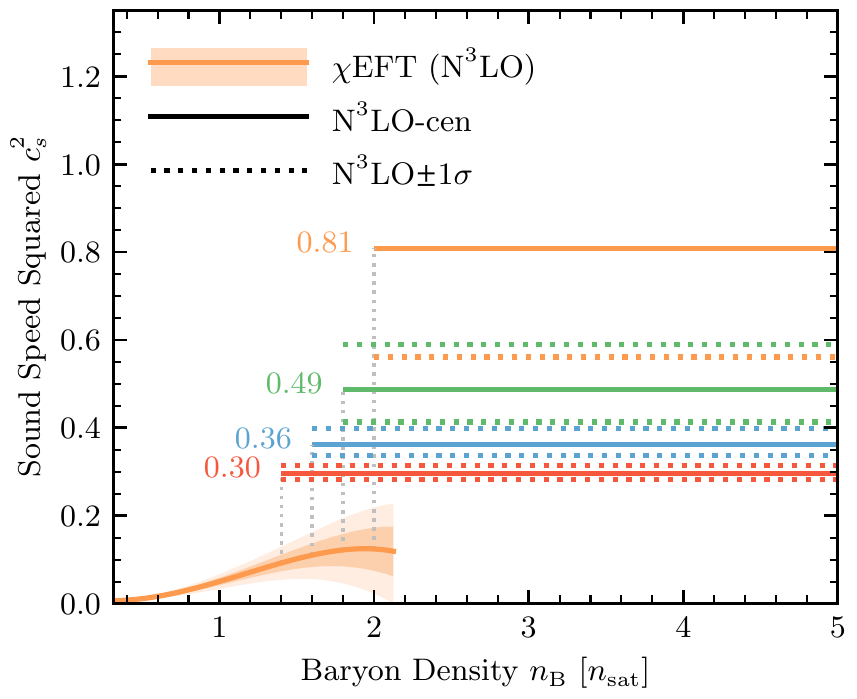}
\end{centering}
\caption{
Sound speed profiles $\csq(\nb)$ for $\NNNLO$-NSM EOSs at low densities (Fig.~\ref{fig:pressure_nsm}), and for matched EOSs with different values of $\csqmat$ at higher densities $\nb\geqslant n_c$ corresponding to $\Rtwo=13\km$ in Fig.~\ref{fig:R20_csq-1} (solid for the central value and dashed for $\pm 1\sigma$ uncertainties); see also Table~\ref{tab:csq_nc}.
}
\label{fig:cs2}
\end{figure}

\begin{figure}[tb]
\begin{centering}
\includegraphics[width=0.49\textwidth]{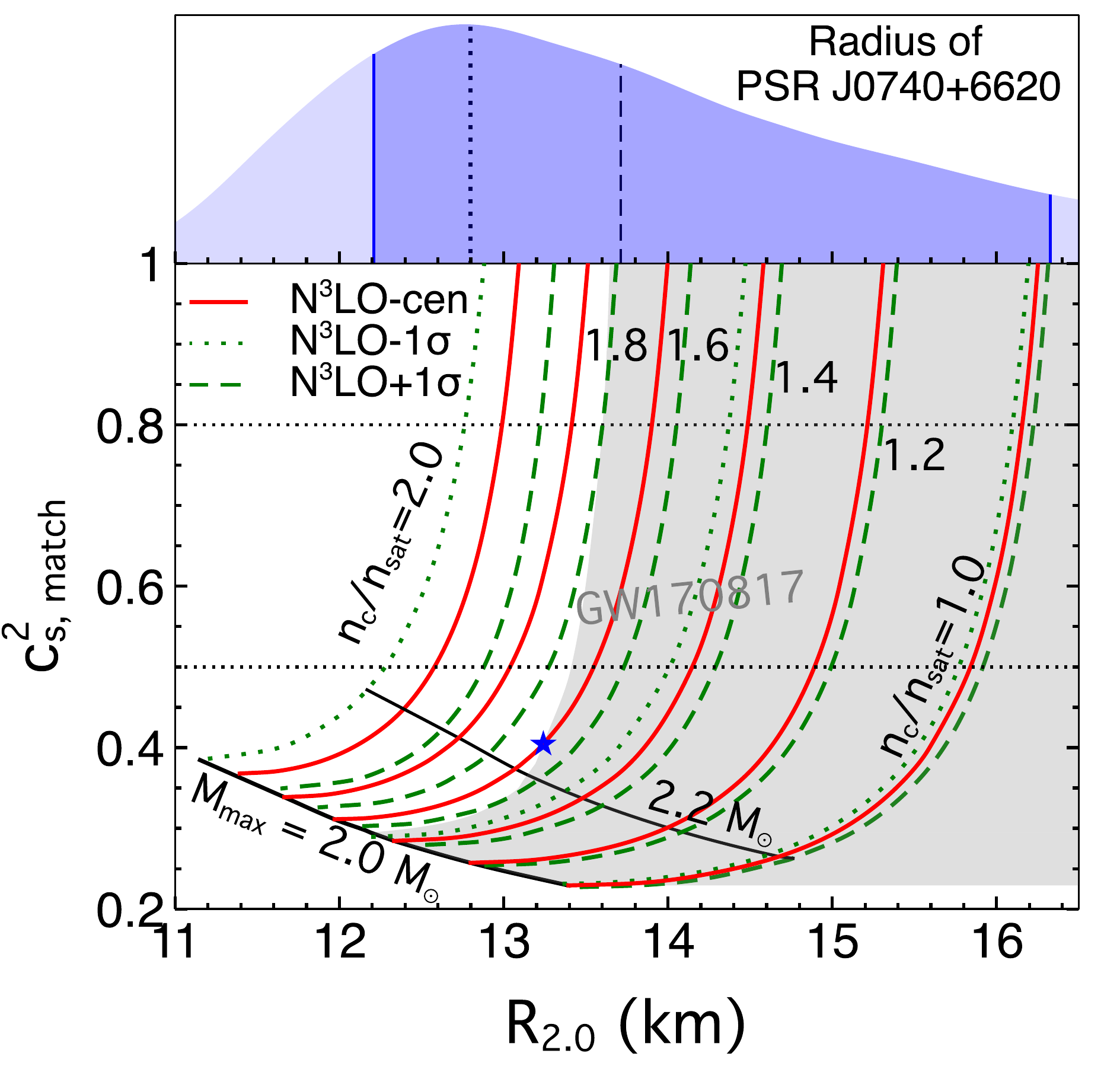}
\end{centering}
\caption{
The top panel shows the joint probability distribution (on an arbitrary scale) for the radius of PSR J0740+6620 as inferred from {\it NICER} and {\it XMM-Newton} data~\cite{Miller:2021qha}. 
The median at $\Rtwo = 13.713 \km$ 
(dashed vertical line) and mode at $\Rtwo = 12.850 \km$ (dotted vertical line) are shown, and the 68\% confidence interval centered around the median is depicted by the dark blue shading, $\Rtwo = (12.209-16.326) \km$. See also Figure~1 and Table~3 in Ref.~\cite{Miller:2021qha}.
The bottom panel: $\Rtwo$--$\csqmat$ plot for $\NNNLO\pm1\sigma$ EOSs in Fig.~\ref{fig:pressure_nsm} up to chosen matching densities in the range $n_c= (1-2)\,\nsat$, assuming there is no additional softening induced by a first-order phase transition at such low densities. The gray-shaded region is excluded by the binary tidal deformability constraint $\tilde\La_{1.186}\leqslant 720$ from GW170817 at the 90\% credibility level~\cite{Abbott:2018wiz} for the \NNNLO-cen EOS. 
The blue asterisk represents intersection between the GW170817 boundary and the $n_c=1.6\,\nsat$ curve for \NNNLO-cen, which corresponds to the maximum $\Rtwo$ consistent with GW constraint if \NNNLO-cen EOS is valid up to $n_c=1.6\,\nsat$ (see discussion in Sec.~\ref{sec:tidal_PT}). 
The two horizontal dotted lines indicate $\csq=0.5$ and $0.8$, respectively. 
}
\label{fig:R20_csq-1}
\end{figure}

Figure~\ref{fig:pressure_nsm} shows the resulting pressure $P(\nb)$ at \NNNLO in neutron star matter as a function of the baryon density $\nb$. The mean value is depicted by the orange solid line, while the dark (light) orange shaded regions correspond to the $1\sigma$ ($2\sigma$) confidence interval. For comparison, we also show the pressures predicted by the phenomenological EOSs NRAPR (green line), SkAPR (blue line), and APR (purple line), as well as microscopic constraints obtained from MBPT~\cite{Hebeler:2013nza} (orange error bar) and Quantum Monte Carlo (QMC) calculations (green error bars)~\cite{Lonardoni:2019ypg}. 
The microscopic calculations used in this work tend to predict somewhat stiffer EOSs compared to other recent microscopic calculations, but still lie on the softer side of the theory-agnostic constraints by Legred~\etal~\cite{Legred:2021hdx} (red error bars).
For instance, in PNM at \NNLO, the EOS considered here predicts $P(2\,\nsat) \approx (20.6 \pm 6.6) \MeV \fmiq$ at the $1\sigma$ confidence level, while state-of-the-art QMC calculations based on a different set of local \chiEFT NN and 3N interactions obtained $P(2\,\nsat) \approx (15.1 \pm 4.7) \MeV \fmiq$ (see Table~2 in Ref.~\cite{Tews:2018kmu}). 
The overall trend that the microscopic calculations used in this work
tend to predict somewhat stiffer EOSs is still present at \NNNLO, although less pronounced, where the MBPT calculations predict $P(2\,\nsat) \approx (18.5 \pm 5.2) \MeV \fmiq$ in PNM.

\begin{table*}[tb]
\caption{
Required stiffening condition (increase in $\csq$) at high densities above $n_c$ for \chiEFT-NSM EOSs (\NNNLO-cen and $\pm 1\sigma$) to reach $\Rtwo=13\km$ (see 
Fig.~\ref{fig:R20_csq-1}); for the full sound speed profiles, see Fig.~\ref{fig:cs2}.
}
\begin{center}
\begin{ruledtabular}
\begin{tabular}{cccc}
$\Rtwo=13\km$ & \multicolumn{3}{c}{$\left[\csqmat, \, \csq(\nb=n_c), \, \De\csq(\nb\geqslant n_c)\right]$} \\[0.5ex]
\hline \\[-2ex]
$n_c/\nsat$ & $-1\sigma$ & \NNNLO-cen & $+1\sigma$  \\[0.5ex]
\hline \\[-2ex]
2.0   & acausal\footnote{for the \NNNLO-$1\sigma$ EOS, the maximum $\Rtwo$ is $12.88\km$ (by setting $\csqmat=1$).} & [0.809, 0.124, 0.685] & [0.562, 0.173, 0.389] \\[0.5ex]
\hline \\[-2ex]
1.8    & [0.589, 0.084, 0.505] & [0.487, 0.122,  0.365] & [0.413, 0.158,  0.255] \\[0.5ex]
\hline \\[-2ex]
1.6  & [0.398, 0.083, 0.315] & [0.364, 0.110,  0.254] & [0.335, 0.137, 0.198] \\[0.5ex]
\hline \\[-2ex]
1.4\footnote{incompatible with GW170817 in the absence of a phase transition.} & [0.314, 0.073, 0.241] & [0.296, 0.092,  0.204] & [0.283, 0.111, 0.172]  \\[0.5ex]
\end{tabular}
\end{ruledtabular}
\end{center}

\label{tab:csq_nc}
\end{table*} 

\section{Speed of sound in the inner core} 
\label{sec:inner_core}

\subsection{Limits on the high-density speed of sound from $\Mmax$ and $\Rtwo$ considerations}
\label{sec:csq_R20}

At higher densities, $\nb > n_c$, we parametrize the inner core using a maximally stiff EOS with a constant sound speed $\csqmat \leqslant 1$, which is stiffer than all other EOSs that have their maximum $\csq$ below $\csqmat$, and when matched smoothly to the outer core, it leads to both the largest neutron star radii and the highest $\Mmax$ (see Ref.~\cite{Drischler:2020fvz} for details). Consequently, $\csqmat$ represents the smallest possible $\csqmax$ for any realistic high-density EOSs to achieve a specific value of neutron star radius and/or $\Mmax$.

We solve the Tolman--Oppenheimer--Volkoff (TOV) equations~\cite{Tolman:1939jz, Oppenheimer:1939ne} of hydrostatic equilibrium for non-rotating neutron stars for a set of EOSs with different $(\csqmat, n_c)$ to map the minimum speed of sound in the core $\mincsq =\csqmat(\Rtwo, n_c)$ needed to support a neutron star with a given radius $\Rtwo$ for matching density $n_c$. 
Since rotational effects alter the equatorial radius only when the neutron star spin frequency $\nu\simeq\nu_{\rm K}$, where $\nu_{\rm K} \approx 1076 \Hz \left(\Rtwo /12 \km \right)^{-3/2}$ is the Keplerian frequency~\cite{Lattimer:2004pg,Haensel:2009wa}, we neglect the effect of rotation on neutron star structure given that the observed spin frequency of PSR J0740+6620 $\nu = 346.5 \Hz$ is small. 

\begin{figure*}[tb]
\parbox{0.5\hsize}{
\includegraphics[width=\hsize]{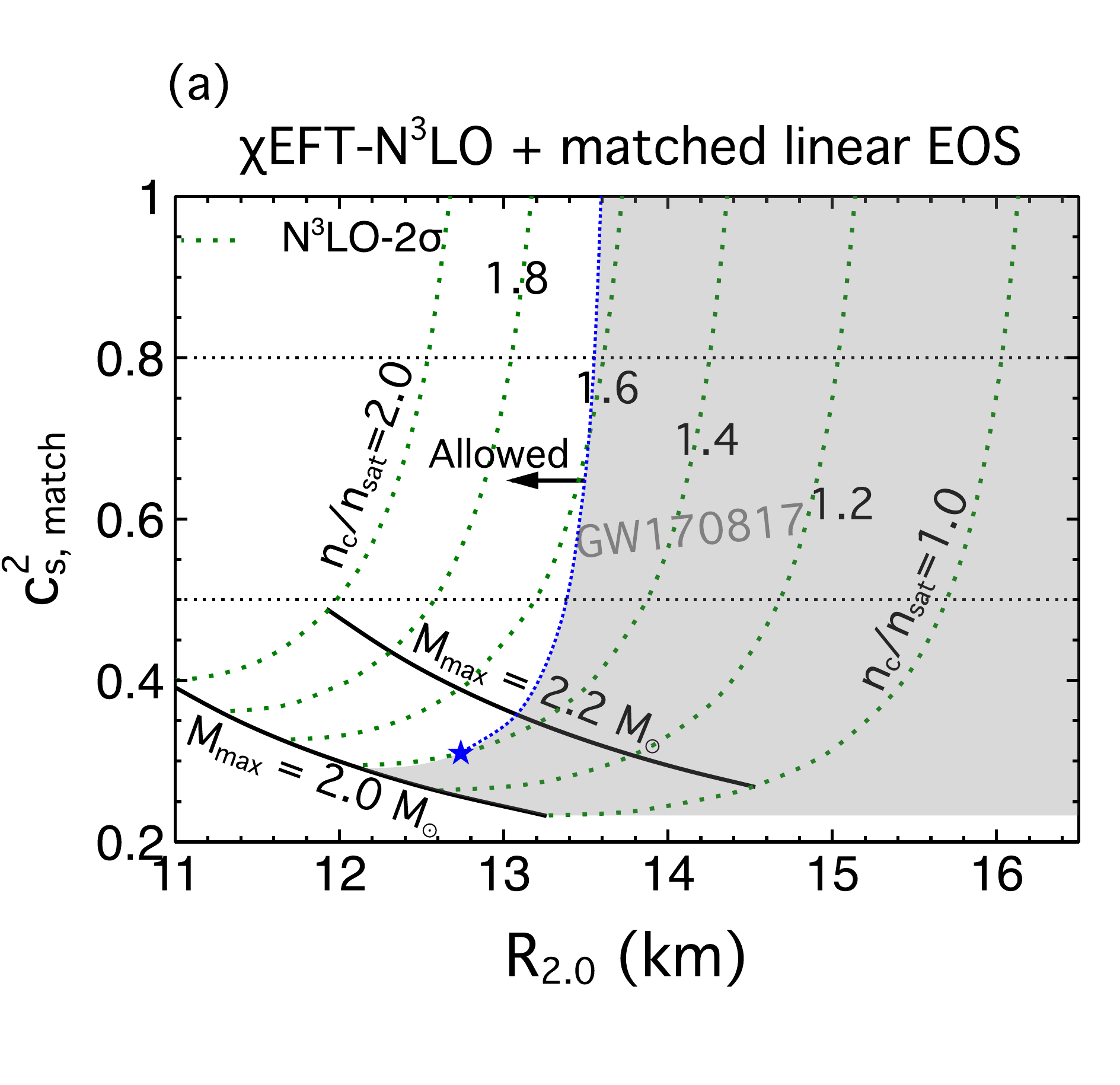}\\[-3ex]
}\parbox{0.5\hsize}{
\includegraphics[width=\hsize]{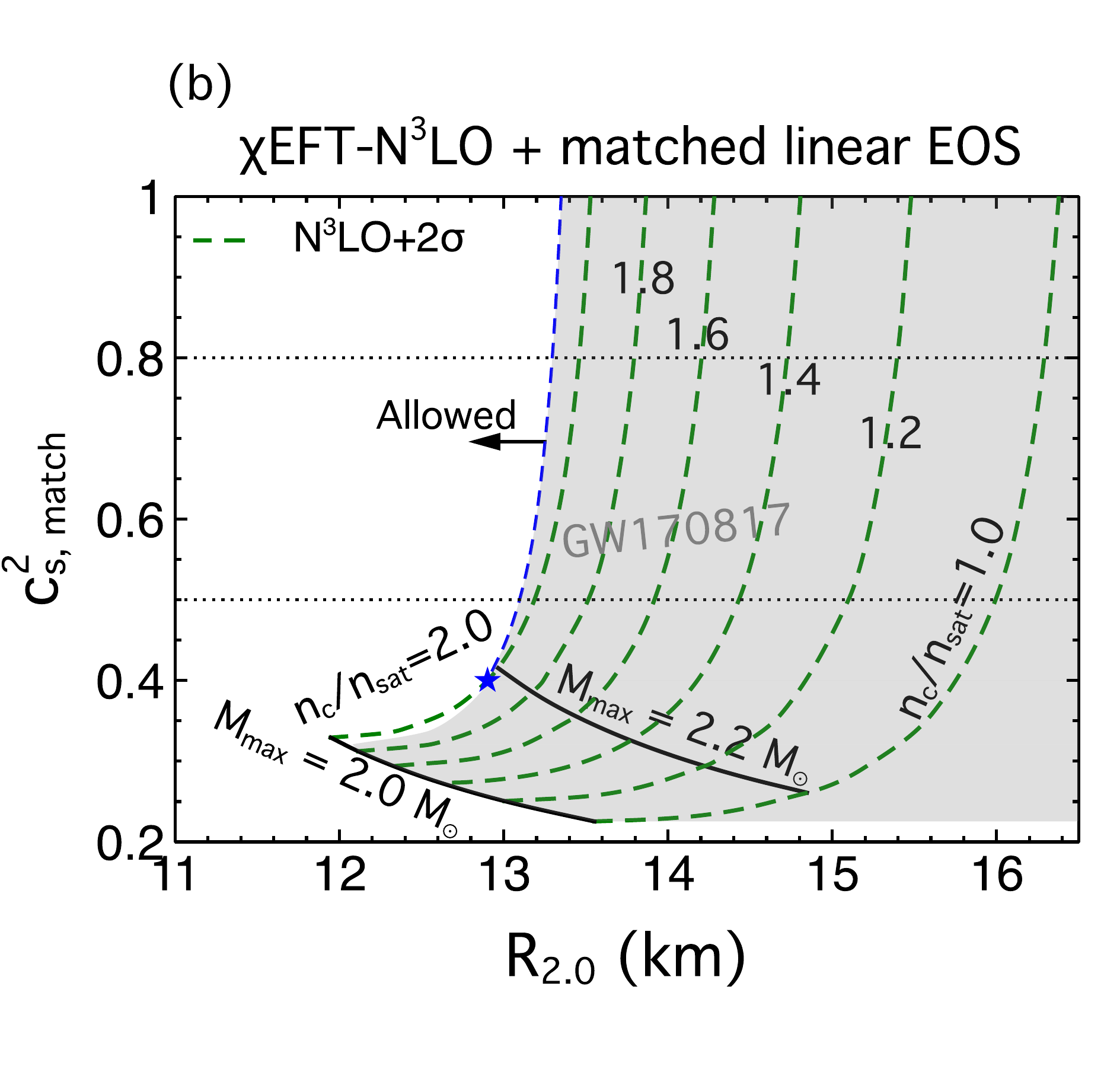}\\[-3ex]
}
\caption{
Similar to Fig.~\ref{fig:R20_csq-1} but for the $\NNNLO\pm2\sigma$ EOSs applied at low densities $1-2\,\nsat$. 
Examples of the revised $\Rtwo$--$\csqmat$ relation 
(and the $\mincsq$ determined thereafter) permitting phase transitions $\De\emat\geqslant 0$ include the blue dotted curve in panel~(a) for $n_c=1.4\,\nsat$ with \NNNLO-2$\sigma$, and the blue dashed curve in panel~(b) for $n_c=2.0\,\nsat$ with \NNNLO+2$\sigma$. 
The blue asterisks indicate where the strength of the phase transition decreases to zero, \ie, no finite discontinuities in the EOS, and the revised $n_c$ curve (blue dotted and blue dashed) is smoothly joined to the part of the original $n_c$ curve that lies outside the gray region (green dotted and green dashed). 
}
\label{fig:R20_csq-2}
\end{figure*}

Figure~\ref{fig:R20_csq-1} shows our central results for \chiEFT-NSM EOS at \NNNLO with $\pm1\sigma$ uncertainties (see also Fig.~\ref{fig:pressure_nsm}), 
supplemented with the radius measurement of PSR J0740+6620 reported by Miller~\etal~\cite{Miller:2021qha} in the top panel.
\footnote{We show the radius distribution function from Miller~\etal~\cite{Miller:2021qha} because it is based on a more agnostic prior (allowing neutron star radii $>16\km$) than the one in Ref.~\cite{Riley:2021pdl}. This choice does not affect our calculations or the derived lower bounds on $\csqmax$.} 
As examples, select combinations of $n_c$ and $\csqmat$ that lead to $\Rtwo=13\km$ are given in Fig.~\ref{fig:cs2} and Table~\ref{tab:csq_nc}, and results for \chiEFT-\NNNLO $\pm2\sigma$ are shown in Fig.~\ref{fig:R20_csq-2}. 
We choose the different matching densities in the range $n_c=(1-2)\,\nsat$ optimized for large values of $\Rtwo$ ($n_c>2.0\,\nsat$ gives rise to even smaller $\Rtwo$).

We find that $\mincsq$ rises more rapidly with $\Rtwo$ if the matching density $n_c$ is fixed at a higher value. 
Assuming the central \chiEFT-\NNNLO EOS (denoted as \NNNLO-cen hereafter) is valid up to 
$n_c=2.0\,\nsat \, (1.8\,\nsat)$, represented by the two leftmost solid red curves in Fig.~\ref{fig:R20_csq-1}, an increase in 
$\Rtwo$ from $12.6 \km$ to $13.1 \km$ would require $\mincsq$ to increase from 
$\approx 0.5 \, (0.39)$ to $\approx 1.0 \, (0.53)$. 
In consequence, for large values of $\Rtwo$ the required increase in $\csq$ (\ie, the rapid stiffening of the EOS) over the density range close to $n_c$ is substantial. 
This is evident from Fig.~\ref{fig:cs2} and Table~\ref{tab:csq_nc}, where we demonstrate the variance in $\csq$ for EOSs that correspond to $\Rtwo=13\km$. Although the situation is ameliorated for the stiffer $+1\sigma$ EOS and for smaller values of $n_c$, magnitudes of the jump $\De\csq$ indicate that above $n_c$ an unusual stiffening in the EOS compared to what \chiEFT predicts towards high density is  necessary, in particular given the fact that $\csq$ rises at a much smaller rate when approaching $\approx 2\,\nsat$ from below.

For a given $n_c$, the stiffer +$1\sigma$ and +$2\sigma$ EOSs (green dashed curves in Figs.~\ref{fig:R20_csq-1} and \ref{fig:R20_csq-2}) result in smaller values of $\mincsq$, and therefore provide conservative estimates about the lower bound on the maximum core speed of sound. 
Considering the stiffest EOS compatible with \chiEFT-\NNNLO at the $2\sigma$ level (see the right panel in Fig.~\ref{fig:R20_csq-2}), the largest possible $\Rtwo$ limited by causality is then 
$\approx 13.53 \km \, (13.87 \km)$ for $n_c=2.0\,\nsat \, (1.8\,\nsat)$ (where the green dashed curves intersect with the upper y-axis), and a measurement of 
$\Rtwo\geqslant13 \km$ indicates $\mincsq\geqslant 0.42\,(0.35)$, which violates the conformal bound $\csq\leqslant1/3$~\cite{Cherman:2009tw,Bedaque:2014sqa,Tews:2018kmu,Landry:2020vaw}. 
Note that at this stage we have not yet taken into account tidal deformability constraints from GW170817~\cite{Abbott:2018wiz} (represented by the gray regions that are excluded) which dominate the upper limit on the stiffness or pressure of EOS at intermediate densities $(2-3)\,\nsat$~\cite{Landry:2020vaw,Legred:2021hdx}. 

Our conclusion that a high sound speed is required in PSR J0740+6620's inner core, if neutron star radii $\gtrsim 13 \km$ are realistic, is a direct consequence of the modest pressures predicted at $\nb \approx (1-2) \, \nsat$. This is not unique to the EOS used here. As indicated in Fig.~\ref{fig:pressure_nsm}, most microscopic calculations to-date find similar soft neutron-rich matter EOSs.
Within the uncertainties, the pressures in neutron-star matter predicted by \chiEFT calculations are not increasing rapidly enough at $\approx 2\,\nsat$. But the microscopic EOS used in this work is stiffer than other EOS constraints from \chiEFT. Hence, our constraints on the minimum core sound speed (at \NNNLO) are conservative, as other microscopic calculations would predict a higher lower bound.

\begin{figure*}[tb]
\parbox{0.5\hsize}{
\includegraphics[width=\hsize]{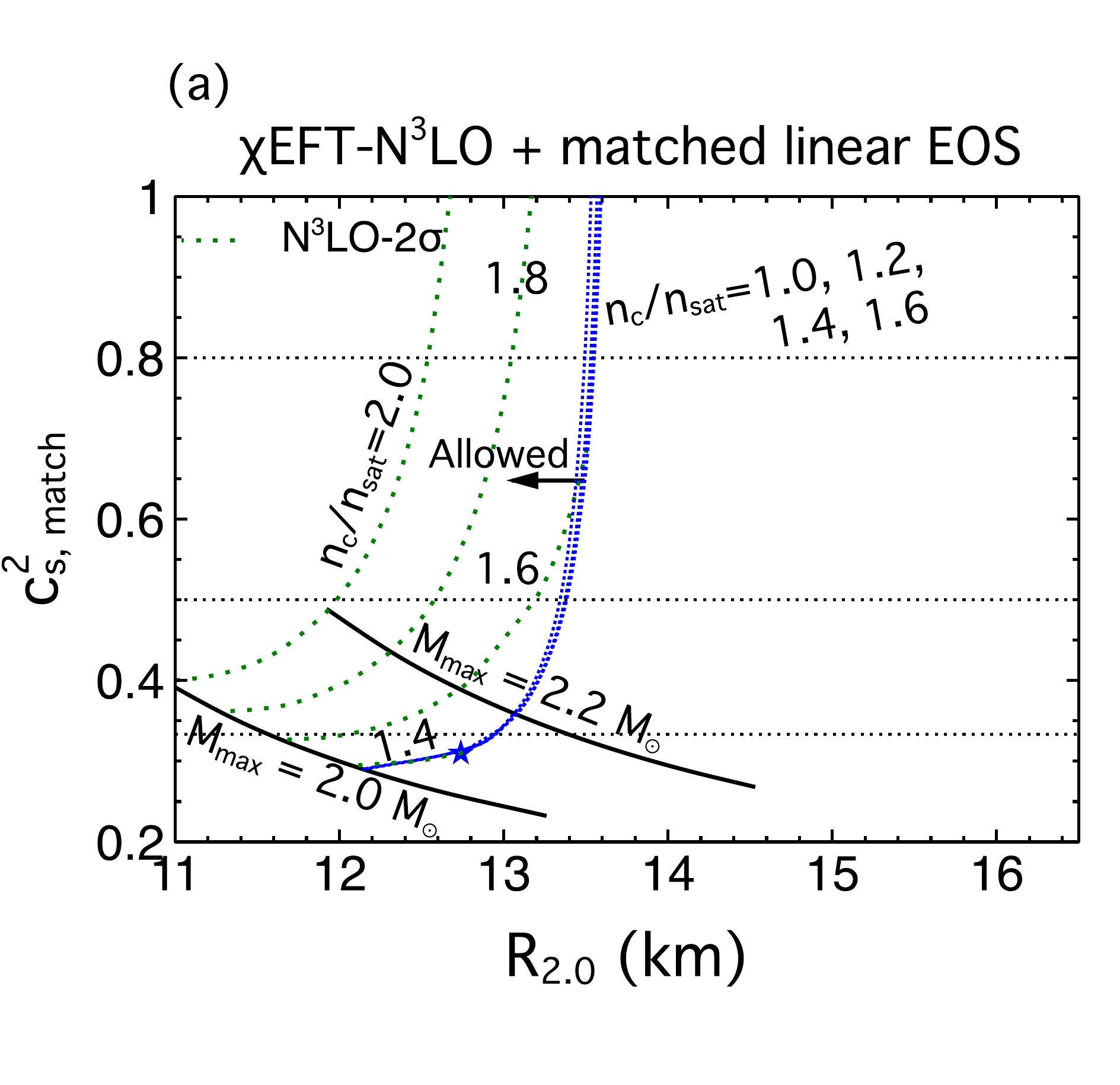}\\[-3ex]
}\parbox{0.5\hsize}{
\includegraphics[width=\hsize]{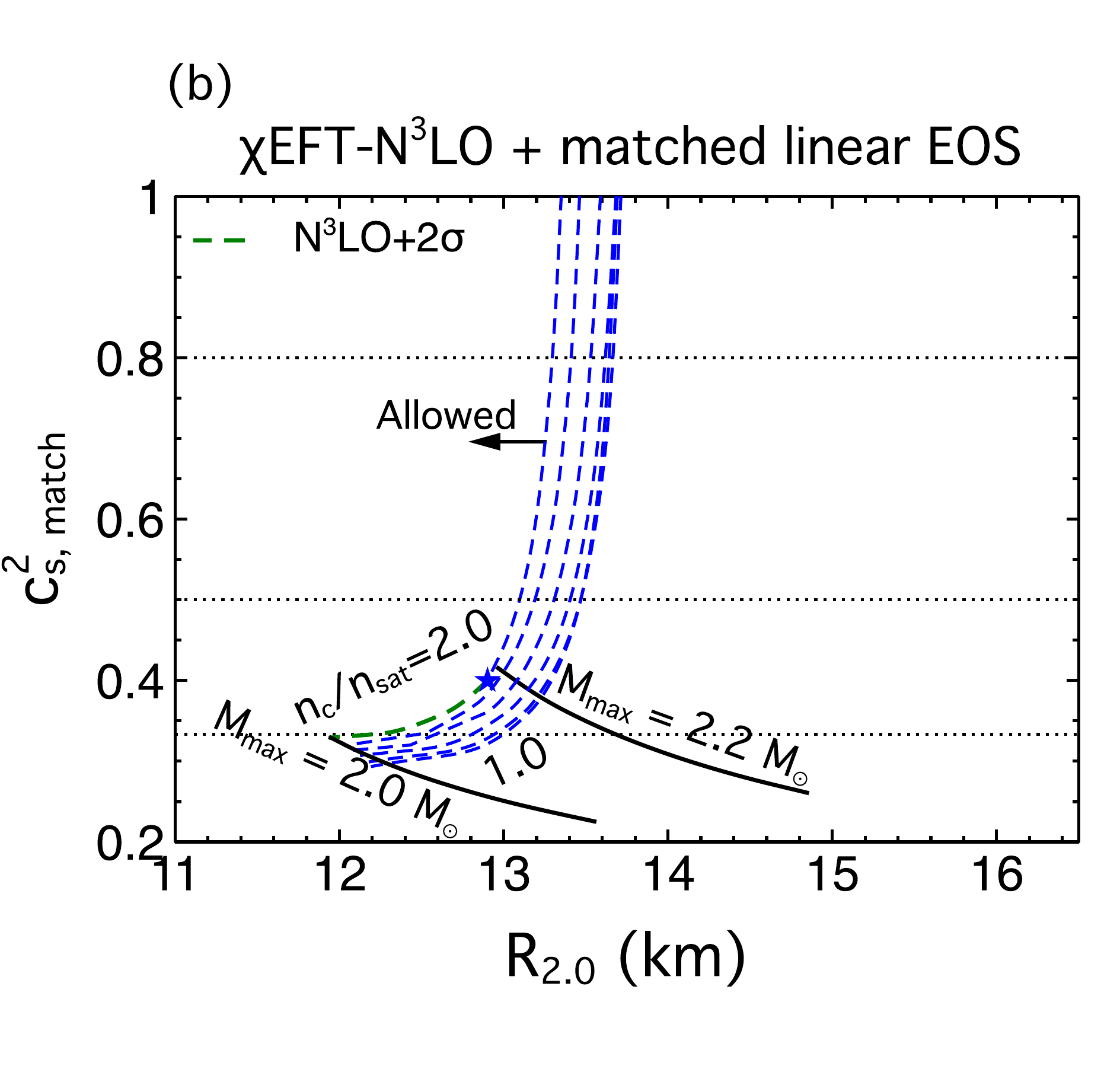}\\[-3ex]
}
\caption{ 
Revised GW170817 boundaries for the $\pm 2\sigma$ EOSs with various values of $n_c$ ($n_c=1.0$, $1.2$, $1.4$, and $1.6\,\nsat$ in panel~(a) and $n_c=1.0$, $1.2$, $1.4$, $1.6$, $1.8$, and  $2.0\,\nsat$ in panel~(b), from right to left) in the presence of a phase transition, represented by the blue dotted and blue dashed curves, which step into the gray excluded regions shown in Fig.~\ref{fig:R20_csq-2}. 
For such low matching densities (or early stiffening in the EOS) and accordingly large $\Rtwo$ to be compatible with ${\tilde\La}_{1.186}\leqslant 720$, a phase transition is needed and the inferred $\mincsq$ for a given $\Rtwo$ is higher because of the (local) extra softening induced at the transition. 
For comparison, the conformal limit $\csq=1/3$ is also indicated (bottom dotted horizontal line).
}
\label{fig:R20_csq-2_PT}
\end{figure*}

There also exists a generic lower bound on $\mincsq$ determined by requiring that the maximum mass of neutron stars is at least $\approx 2\,\Msun$, as indicated in Figs.~\ref{fig:R20_csq-1} and~\ref{fig:R20_csq-2}  by the black solid boundaries where all colored curves end. Here, we choose $\Mmax=2.0\,\Msun$ and $2.2\,\Msun$ to display the trend.  
A smaller $\Mmax$ is accompanied by a smaller $\Rtwo$, and for any assumed lower bound on $\Mmax$, the corresponding lower bound on $\mincsq$ increases when $n_c$ increases, moving from the right to the left along the black solid boundaries.

\subsection{Accommodating GW170817 constraints and the role of phase transitions}
\label{sec:tidal_PT}

In this section, we focus on the compatibility between small tidal deformability constraints from GW170817 and possible (large) radius constraints on massive neutron stars. Using the standard PhenomPNRT model, the binary chirp mass $\Mchirp=1.186\pm0.001\,\Msun$ and the binary tidal deformability ${\tilde\La}_{1.186}\leqslant720$ (90\% credibility level) of GW170817 were obtained by gravitational waveform fitting~\cite{Dietrich:2018uni,Abbott:2018wiz}. 
We briefly describe how the upper bound on $\tilde\La$ translates into constraints on the high-density EOS parameters and effectively the inferred $\Rtwo$. We also investigate the modification to the minimally required core sound speed $\mincsq$ when GW170817 is taken into account, with or without sharp phase transitions, \ie, finite discontinuities in the energy density $\De\ep$ in the EOS. 

Figure~\ref{fig:R20_csq-1} shows the gray-shaded region on the $\Rtwo$--$\csqmat$ plane, which is excluded by GW170817 (${\tilde\La}_{1.186}\leqslant720$, \NNNLO-cen EOS), assuming that no phase transition occurs at $\nb=(1-2)\,\nsat$. It is clear that low matching densities $n_c\lesssim 1.5\,\nsat$ disagree with the GW data; for a given $n_c$, the largest possible $\Rtwo$ consistent with small tidal deformabilities observed is therefore determined by the intersection between the GW170817 boundary and the corresponding $n_c$ contour. 
For $n_c=1.6\,\nsat$, this value is $\Rtwo\approx13.24 \km$ (blue asterisk) alongside a high speed of sound in the core $\csqmat\approx 0.404$. Tracking the GW170817 boundary upwards, both $n_c$ and $\Rtwo$ grow gradually with a steep increase in $\csqmat$.

If the low-density EOS is softer, it is relatively simple to meet the GW170817 constraints, which allows more space for matching at smaller $n_c$; see the gray region in Fig.~\ref{fig:R20_csq-2}~(a) for the \NNNLO-$2\sigma$ EOS compared to that in  Fig.~\ref{fig:R20_csq-1}. On the other hand, for the stiffest \NNNLO+$2\sigma$ EOS, the tension with the GW170817 constraints is more severe and  matching densities $\lesssim2.0\,\nsat$ are almost entirely ruled out (see Fig.~\ref{fig:R20_csq-2}~(b)).

\begin{figure*}[tb]
\parbox{0.5\hsize}{
\includegraphics[width=\hsize]{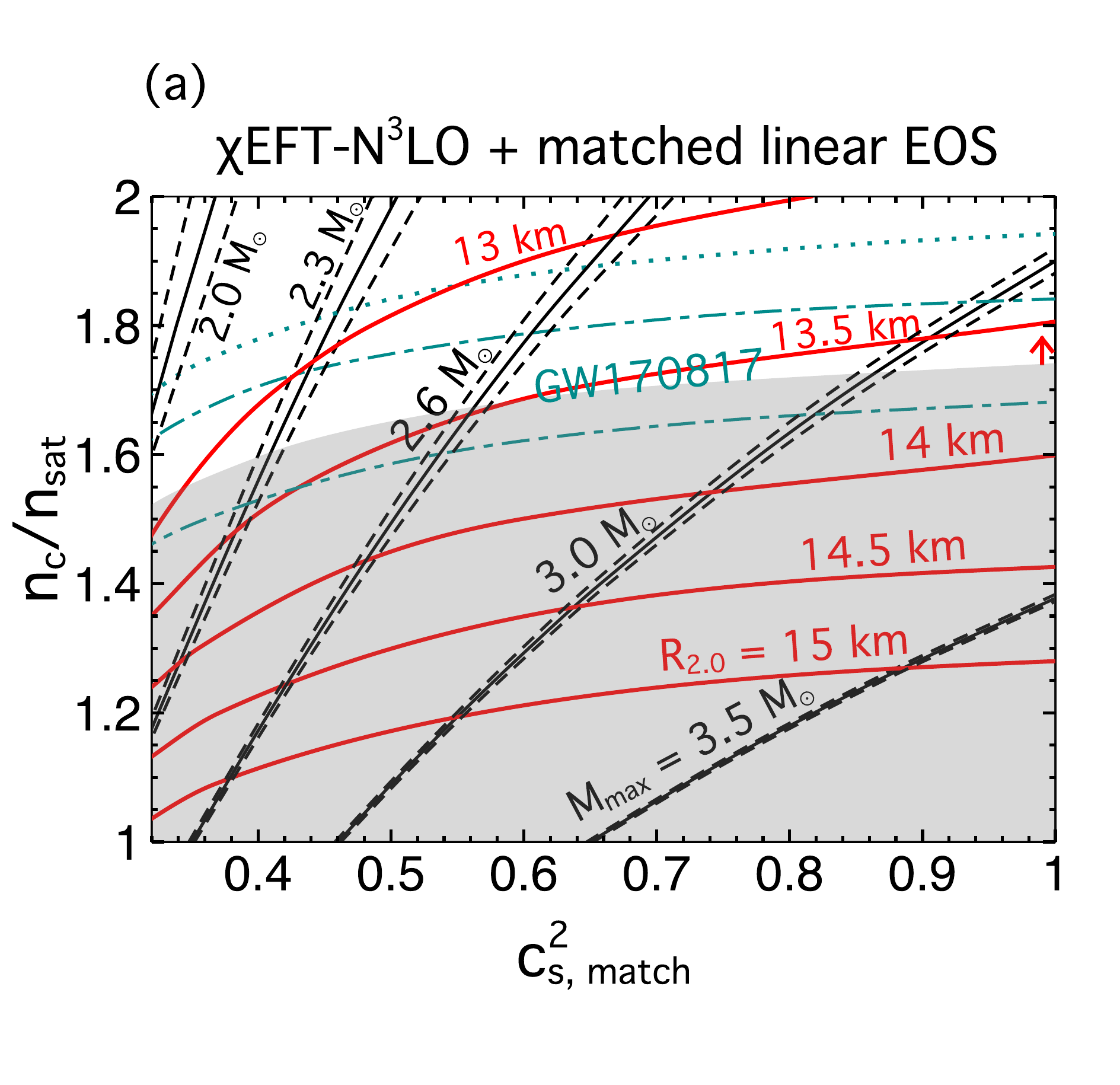}\\[-3ex]
}\parbox{0.5\hsize}{
\includegraphics[width=\hsize]{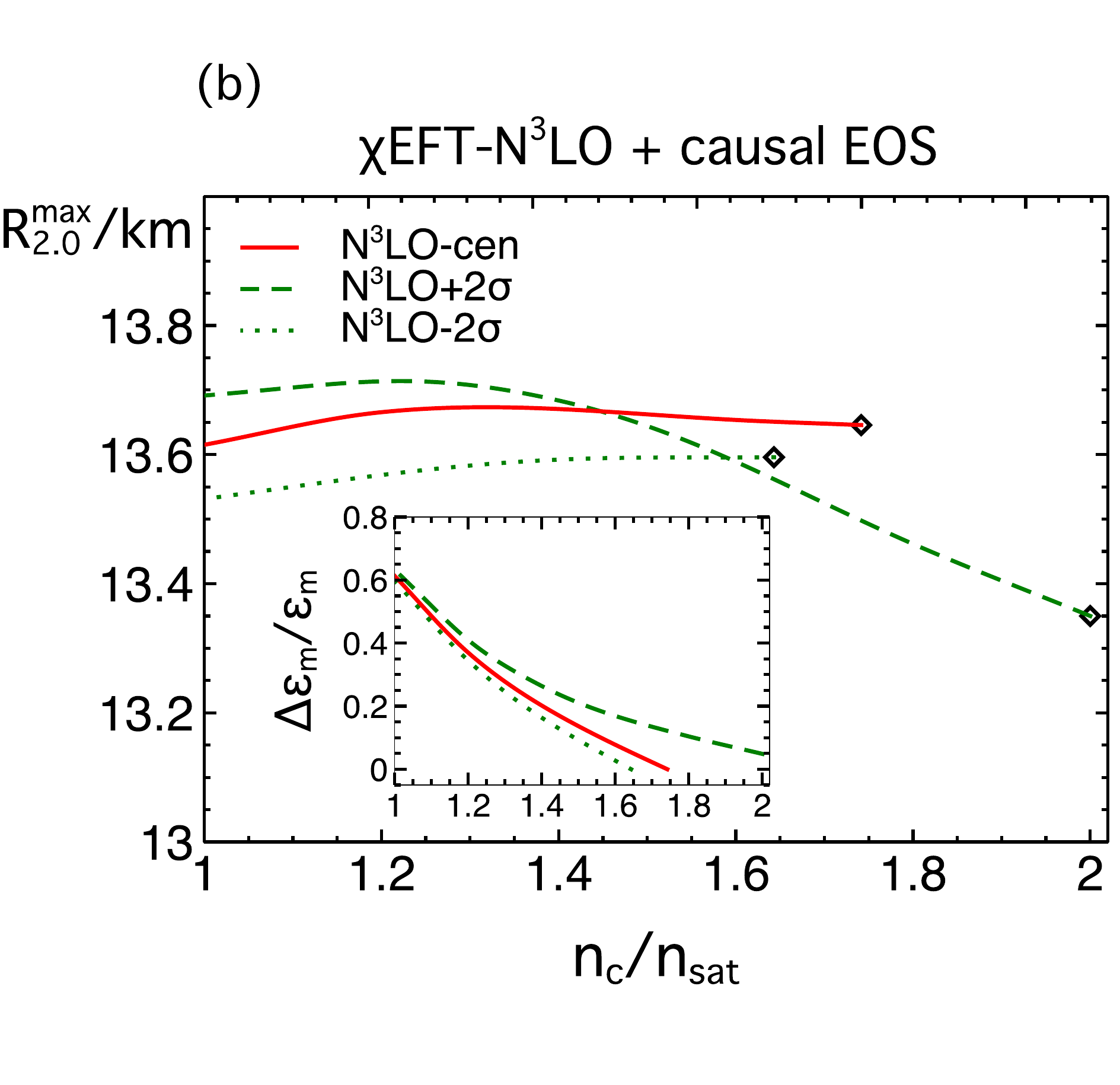}\\[-3ex]
}
\caption{Panel~(a): $\Rtwo$ and $\Mmax$ contours on the $(\csqmat, n_c)$ plane, obtained without a phase transition in the density interval $\nb=(1-2)\,\nsat$. 
For each value of $\Mmax$, the central solid curves represent results with \NNNLO-cen; dashed lines indicate $\pm1\sigma$ bounds. For $\Rtwo$ contours, only results with \NNNLO-cen are shown (red solid lines). The gray-shaded region is excluded by the binary tidal deformability constraint $\tilde\La_{1.186}\leqslant 720$ from GW170817 at the 90\% credibility level~\cite{Abbott:2018wiz} if \NNNLO-cen is assumed; the (cyan) dot-dashed lines refer to constraints with the $\pm 1\sigma$ uncertainties. 
The thin dotted line indicates an even lower upper bound with \NNNLO-cen and $\tilde\La_{1.186}\leqslant 600$. 
Panel~(b): maximally allowed value of $\Rtwo$ that is compatible with $\tilde\La_{1.186}\leqslant 720$ by matching to the causal EOS with a density discontinuity $\De\emat/\emat\geqslant 0$ (shown in the inset) at $n_c=(1-2)\,\nsat$, below which the \NNNLO-cen and $\pm 2\sigma$ EOSs are used. The value of $\Rtwo^{\rm max}$ associated with the highest $n_c$ on each curve, represented by the open diamonds, corresponds to the intersection points between the upper y-axis and the boundary of the gray region in Figs.~\ref{fig:R20_csq-1} and~\ref{fig:R20_csq-2}. Note, however, that these extremely large $\Rtwo$ are also indicative of very high maximum masses $\gtrsim 2.8\,\Msun$ (see panel~(a) for an estimate), above the secondary component mass of GW190814~\cite{LIGOScientific:2020zkf}.
}
\label{fig:R20_Mmax-N3LO}
\end{figure*}

We turn now to the discussion of incorporating effects from sharp phase transitions. 
For instance in Fig.~\ref{fig:R20_csq-2}~(a), the original $n_c=1.4\,\nsat$ curve (green dotted lines) without additional softening in the EOS mostly lies inside the gray region (except the short piece below the blue asterisk), which is incompatible with GW170817, \ie, violating ${\tilde\La}_{1.186}\leqslant720$, 
and such small values of $n_c$ should be considered nearly excluded. However, by introducing a finite density discontinuity $\De\emat$ at $\nb=n_c$, smaller tidal deformabilities at masses relevant to GW170817 become feasible, and the resulting $\csqmat$--$\Rtwo$ relation for $n_c=1.4 \,\nsat$ (that allows a phase transition) determined by the limiting case ${\tilde\La}_{1.186}=720$ is shifted toward the left (blue dotted lines), compared to the previous curve that lies inside the gray excluded region (green dotted lines).  
Interestingly, this updated $n_c=1.4\,\nsat$ curve (blue dotted line) closely tracks the original GW170817 boundary, the edge of the gray region, which assumes no phase transition for all $n_c$'s.

To further explore modifications that sharp phase transitions at different matching densities brought about to the GW170817 boundary, we display in Fig.~\ref{fig:R20_csq-2_PT} the associated curves for which $n_c\leqslant1.6\,\nsat$ (panel~(a), \NNNLO-2$\sigma$ EOS) and $n_c\leqslant2.0\,\nsat$ (panel~(b), \NNNLO+2$\sigma$ EOS), respectively. 
We find that smaller $n_c$ allows slightly larger $\Rtwo$ with the revised boundary moving toward the right in the $\Rtwo$--$\csqmat$ plane.  
It is obvious that a finite density discontinuity $\De\emat$ at the phase transition is effective in accommodating GW170817 owing to the preferable softening, but, correspondingly, the necessity for increasing $\csqmax$ at higher densities is inevitably strengthened, leading to sizeable corrections to the $\mincsq$ indicated from the previous green curves inside the gray regions of Fig.~\ref{fig:R20_csq-2} (which assumed no phase transition). 

In essence, introducing sharp phase transitions in the EOS between $\nb=(1-2)\,\nsat$ has three principal outcomes: i)~it is now possible for small matching densities $n_c\lesssim 1.4-1.6\,\nsat$ to be compatible with GW170817 provided that the corresponding discontinuity in energy density is sufficiently large; ii)~depending on the specified value of $n_c$, the binary tidal deformability upper bound $\tilde\La_{1.186}= 720$ now translates into different GW170817 boundaries (see Fig.~\ref{fig:R20_csq-2_PT}) on the $\Rtwo$--$\csqmat$ plane; iii)~the lower bound on the core sound speed $\mincsq$ (to satisfy large $\Mmax$ and/or large $\Rtwo$) increases due to the added softening of EOS at low densities.

Our results indicate that to satisfy both small tidal deformabilities of GW170817 and $\Rtwo\geqslant 13 \km$, the conservative lower bounds [see Fig.~\ref{fig:R20_csq-2_PT}~(b)] on the minimum core speed of sound should be $\mincsq\geqslant 0.442\,(0.411)$ and $\Rtwo^{\rm max}\approx 13.35 \km \, (13.46 \km)$, if \chiEFT-\NNNLO is assumed valid up to 
$n_c=2.0\,\nsat \, (1.8\,\nsat)$. 
In contrast, the values previously inferred without taking into account GW170817 constraints were 
$\mincsq\geqslant 0.42\,(0.35)$ and $\Rtwo^{\rm max}\approx 13.53\km \, (13.87 \km)$, respectively (see discussion in Sec.~\ref{sec:csq_R20} and Fig.~\ref{fig:R20_csq-2}~(b)). 
Similarly, for a smaller lower bound $\Rtwo\geqslant 12.5 \km$, Fig.~\ref{fig:R20_csq-2_PT}~(b) tells that $\mincsq\geqslant 0.348\,(0.333)$ for $n_c=2.0\,\nsat \, (1.8\,\nsat)$.

It is worth noting that, although there is still a chance for the conformal limit $\mincsq\leqslant 1/3$ to be compatible with $\Mmax\geqslant 2.0\,\Msun$ and GW170817 (the small triangle region in Fig.~\ref{fig:R20_csq-2_PT}~(a) and~(b) encompassed by the rightmost blue curve, bottom black solid curve, and bottom horizontal dotted line), the permitted range of $\Rtwo$ is quite narrow, \ie, $\Rtwo = (11.55-13.00)\km$ for \chiEFT-\NNNLO 
at the $2\sigma$ confidence level. 
Moreover, a slight increase in $\Mmax\gtrsim 2.1\,\Msun$ would suffice to exclude $\mincsq\leqslant 1/3$~\cite{Drischler:2020fvz}.

Figure~\ref{fig:R20_Mmax-N3LO}~(b) shows the absolute upper bounds on $\Rtwo$ imposed by causality and GW170817 with varying $n_c$. 
We find that $\Rtwo^{\rm max}$ is relatively insensitive to $n_c$, except for the stiffest \NNNLO+$2\sigma$ EOS. 
The inset depicts the corresponding phase transition strengths $\De\emat/\emat$ at different matching densities, where $\emat$ is the energy density in the \chiEFT EOS at $\nb=n_c$. Values of $\Rtwo^{\rm max}$ reached at the highest $n_c$ (and simultaneously the smallest $\De\emat/\emat$) are denoted by the open diamonds, which also correspond to where the upper y-axis intersects with the boundary of the gray regions in Figs.~\ref{fig:R20_csq-1} and~\ref{fig:R20_csq-2}.
It is reasonably justified to claim that despite the presence or absence of unusual softening in the EOS at low densities $(1-2)\,\nsat$, the maximal achievable radius for a $2\,\Msun$ neutron star is 
$\Rtwo^{\rm max}\approx 13.7 \km$ limited by causality and GW170817. Note that this is close to the central value reported in Ref.~\cite{Miller:2021qha} (see also the top panel in Fig.~\ref{fig:R20_csq-1}); should future observations confirm $\Rtwo>13.7\km$, it would strongly favor a transition at sub-saturation density to an EOS that is significantly stiffer than predicted by \chiEFT calculations.

We also show in Fig.~\ref{fig:R20_Mmax-N3LO}~(a) how both $\Mmax$ and $\Rtwo$ depend on $n_c$ (chosen to lie within $1-2\,\nsat$) and $\csqmat$ at high densities, assuming the validity of \chiEFT-\NNNLO up to $n_c$ and $\De\emat=0$. 
Because finite $\De\emat$ would only reduce both $\Mmax$ and $\Rtwo$, the values of $\csqmat$ indicated from  Fig.~\ref{fig:R20_Mmax-N3LO}~(a) by imposing lower bounds on $\Mmax$ or $\Rtwo$ serve as conservative estimates for $\mincsq$. 
We find that $\mincsq$ is in general more sensitive to $\Rtwo$ than to $\Mmax$: 
an increase in $\Rtwo$ from $13.0\,\km$ to $13.5\,\km$ is more constraining than an increase in $\Mmax$ from 
$2.0\,\Msun$ to $2.3\,\Msun$, pushing $\mincsq$, \eg, from 
$\approx0.46$ to $\approx0.80$ in the former case compared to 
$\approx0.33$ to $\approx0.45$ in the latter case for $n_c=1.75\,\nsat$.

Despite different inference methods used, recent works applying {\it NICER} measurements of PSR J0740+6620 to extract neutron star properties and EOS constraints obtained results that are broadly consistent~\cite{Raaijmakers:2021uju,Pang:2021jta,Legred:2021hdx,Biswas:2021yge}. 
The \NNNLO+$2\sigma$ EOS in the outer core used in this work is relatively stiff at $\nb \gtrsim \nsat$ compared to the \chiEFT models applied in Ref.~\cite{Raaijmakers:2021uju,Pang:2021jta} (see also Fig.~6 in Ref.~\cite{Raaijmakers:2021uju} for a comparison), and its pressure coincides with the mean value inferred from theory-agnostic study in Ref.~\cite{Legred:2021hdx} at $2\,\nsat$ as shown in Fig.~\ref{fig:pressure_nsm}. Hence, we expect our estimates on $\mincsq$ to be conservative. 
Given the still large uncertainties involved, the consensus is that current combined data from GW, x-ray, and radio observations are not yet informative enough to identify or rule out microscopic models that exhibit first-order or crossover transitions into exotic matter at densities relevant for neutron stars (for different phase transition scenarios explored, see, \eg, Refs.~\cite{Somasundaram:2021ljr,Zhao:2020dvu,Han:2020adu,Tang:2021snt,Tan:2021ahl,Li:2021sxb,Christian:2021uhd}). 
We investigated in detail the enhancement in $\mincsq$ to reach large $\Rtwo$ when the EOS undergoes a finite discontinuity $\De\emat$ at low densities, limited by the small tidal deformabilities measured in GW170817.  
The possibility of phase transitions in the density interval $(1-2)\,\nsat$ is of particular interest for experimental probes such as low-to-intermediate heavy-ion collisions~\cite{Danielewicz:2002pu}, as existing analyses of these experiments have largely been done with nucleonic degrees of freedom only.

\section{Summary and Outlook} 
\label{sec:cons}

In this \article, we have addressed how nuclear physics constraints on the EOS of matter at the moderate densities encountered in the neutron star outer core can be combined with recent {\it NICER} observations to provide a lower bound on the maximum speed of sound in dense QCD.
We employed recent \chiEFT calculations of the nuclear EOS up to \NNNLO with EFT truncation errors quantified to explore the implications of the large inferred radius of PSR J0740+6620---the heaviest of the precisely measured two-solar mass neutron stars. We found that the minimum required value for the highest speed of sound reached in the inner core increases rapidly with the radius of massive neutron stars. 
 
If \chiEFT is an efficient expansion for nuclear interactions at $\nb \lesssim 2\,\nsat$, a lower limit on 
$\Rtwo > 12.5 \km$ indicates $\csq \gtrsim 0.36$ in the neutron star inner core, and this lower bound is very sensitive to $\Rtwo$. For 
$\Rtwo =13.1 \km$, we found that $\mincsq=1.0$ when we used the central values for the \chiEFT EOS (see Fig.~\ref{fig:R20_csq-1}), and 
$\mincsq=0.5$ for the stiffest EOS compatible with \chiEFT at the $2\sigma$ level (see Fig.~\ref{fig:R20_csq-2}~(b)). 
Together with the predictions of perturbative QCD at asymptotically high densities ($\gtrsim 40\,\nsat$), this implies that the sound speed must be a non-monotonic function of density, with at least two extrema~\cite{Bedaque:2014sqa,Tews:2018iwm}. If \chiEFT\ constraints are used at densities $\lesssim 1.5\,\nsat$ only, it is possible to accommodate a scenario in which the sound speed is a monotonic function of density. However, in this case, the sound speed must increase very rapidly to its asymptotic value of $c_s=1/\sqrt{3}$ within the neutron star. More significantly, the results in Fig.~\ref{fig:R20_csq-2_PT} indicate that a monotonically increasing sound speed in QCD implies that $\Mmax \lesssim 2.1\,\Msun $ and $\Rtwo \lesssim 13$ km. 

Our conclusion that the large inferred radius of PSR J0740+6620 favors high sound speed is further strengthened by several other calculations of the EOS based on phenomenological models of NN interactions constrained by scattering data and simple models for 3N interactions~\cite{Akmal:1998cf,Pieper:2001ap,Gandolfi:2011xu}. 
Generically, these models predict smaller pressure as well as smaller sound speed values at $\nb \lesssim 2 \,\nsat$ compared to the stiffest EOS compatible with \chiEFT. A common feature shared by these models and \chiEFT is the important role of repulsive 3N forces. 
We find that, even in phenomenological models with strong 3N forces, 
$\Rtwo > 13 \km$ can only be accessed if $\mincsq > 0.5$. 

For EOS calculations based on \chiEFT, it is important to explore EFT truncation errors at high densities for a wide range of potentials as well as the regulator dependence further. 
In particular, a full Bayesian analysis of the nuclear EOS in which also the uncertainties from the low-energy couplings in the nuclear interactions (in addition to the EFT truncation error) are quantified is required.
This will lead to the development of improved order-by-order chiral NN and 3N potential up to \NNNLO with uncertainties rigorously quantified. 
An important step toward this goal has recently been achieved in Ref.~\cite{Wesolowski:2021cni}, where a set of order-by-order chiral NN and 3N interactions with theoretical uncertainties fully quantified has been constructed up to \NNLO. Detailed comparison with the predictions of \chiEFT with $\Delta$ baryons could also provide valuable insights~\cite{Ekstrom:2017koy,Piarulli:2019pfq,Jiang:2020the}. 

Our findings underscore the need to improve constraints on the neutron-rich matter EOS in the density range $\nb \approx (1-2)\,\nsat$. It is fortuitous that both theory and experiment can access these densities in the next 5 to 10 years. Efforts to test and constrain \chiEFT predictions at densities $\nb\simeq \nsat$ using neutron-rich nuclei 
as probes, including improved measurements of neutron-skin thicknesses and dipole polarizabilities, 
would be valuable and are anticipated in the near-term future~\cite{Becker:2018ggl,Budker:2020zer}.
Heavy-ion experiments that study collisions of neutron-rich nuclei at intermediate energy can provide guidance for the behavior of the EOS in the density regime of $(1-2)\,\nsat$~\cite{Danielewicz:2002pu,Estee:2021khi,Huth:2021bsp,FRIB400}. 
When combined with improved astrophysical constraints on the radius of massive neutron stars, these developments can significantly tighten the lower bound on the maximum speed of sound. 
Even a modest reduction in the uncertainty associated with the extraction of the radius of PSR J0740+6620 would have a significant impact. In this context, the \chiEFT-based predictions for viable neutron star radii we discussed in this work could provide useful priors for astrophysical modeling efforts. 

For these reasons, the density interval $\nb \approx (1-2)\,\nsat$ is emerging to be the \textit{golden window} of neutron star physics, in which we can expect significant advances across multiple disciplines soon. 
If these advances strengthen the case for a large sound speed inside massive neutron stars, it would imply the presence of a strongly interacting phase of relativistic matter in their inner cores. 
Such matter would be characterized by a sound speed that increases rapidly over a relatively small density range and would be quite distinct from neutron-rich matter in the outer core, or weakly interacting quark matter.

\begin{acknowledgments}

We would like to thank Constantinos Constantinou, Cole Miller, and Madappa Prakash for sharing their insights with us and are grateful to the National Science Foundation's \textit{Physics Frontier Center: The Network for Neutrinos, Nuclear Astrophysics, and Symmetries} (\href{https://n3as.berkeley.edu}{N3AS}) for encouragement and support. 
This material is based upon work supported by the U.S. Department of Energy, Office of Science, Office of Nuclear Physics, under the FRIB Theory Alliance award DE-SC0013617.
S.H. is supported by the National Science Foundation, Grant PHY-1630782, and the Heising-Simons Foundation, Grant 2017-228. 
The work of S.R. was supported by the U.S. DOE under Grant No. DE-FG02-00ER41132. \\

\end{acknowledgments}


\bibliography{paper}
\end{document}